\documentclass[preprint,review,12pt]{elsarticle}
\usepackage[margin=2.5cm]{geometry}

\usepackage{amssymb}
\usepackage{amsmath}
\usepackage{amsthm}
\usepackage{soul}
\usepackage{color}

\makeatletter
\def\ps@pprintTitle{%
  \let\@oddhead\@empty
  \let\@evenhead\@empty
  \let\@oddfoot\@empty
  \let\@evenfoot\@oddfoot
}
\makeatother

\begin{document}
\begin{frontmatter}

\title{Significantly Enhanced Interfacial Thermal Transport between Single-layer Graphene and Water Through Basal-plane Oxidation}

\author[inst1]{Haoran Cui}
\author[inst1]{Iyyappa Rajan Panneerselvam}
\author[inst4]{Pranay Chakraborty}
\author[inst2]{Qiong Nian}
\author[inst3]{Yiliang Liao}
\author[inst1]{Yan Wang\fnref{myfootnote}}\cortext[mycorrespondingauthor]{Corresponding author}
\ead{yanwang@unr.edu}

\affiliation[inst1]{organization={Department of Mechanical Engineering, University of Nevada, Reno},
            city={Reno},
            postcode={89557}, 
            state={NV},
            country={USA}}
\affiliation[inst2]{organization={School of Engineering for Matter, Transport and Energy, Arizona State University},
            city={Tempe},
            postcode={85281}, 
            state={AZ},
            country={USA}}
  \affiliation[inst3]{organization={Department of Industrial and Manufacturing Systems Engineering, Iowa State University},
            city={Ames},
            postcode={50011}, 
            state={IA},
            country={USA}}  
  \affiliation[inst4]{organization={Department of Engineering and Physics, Southern Arkansas University},
            city={Magnolia},
            postcode={71753}, 
            state={AR},
            country={USA}}  

\begin{abstract}
Heat transfer between graphene and water is pivotal for various applications, including solar-thermal vapor generation and the advanced manufacturing of graphene-based hierarchical structures in solution. In this study, we employ a deep-neural network potential derived from ab initio molecular dynamics to conduct extensive simulations of single-layer graphene-water systems with different levels of oxidation (carbon/oxygen ratio) of the graphene layer. Remarkably, our findings reveal a one-order-of-magnitude enhancement in heat transfer upon oxidizing graphene with hydroxyl or epoxide groups at the graphene surface, underscoring the significant tunability of heat transfer within this system. Given the same oxidation ratio, more dispersed locations of functional groups on graphene surface leads to faster heat dissipation to water.
\end{abstract}

\end{frontmatter}

\section{Introduction}

Graphene is an extraordinary two-dimensional material renowned for its unique physical properties \cite{novoselov2012roadmap}. Single-layer graphene, for instance, exhibits remarkably high in-plane thermal conductivity ($\kappa$) ranging from 1,500 to 5,000 W/m-K when free-standing \cite{balandin2008superior, cai2010thermal}. However, when graphene is supported on a substrate, its $\kappa$ decreases to approximately 600 W/m-K at room temperature \cite{seol2010two}. The phonon-mediated heat conduction in graphene, whether in pristine, defected, or complex architectural forms, has been extensively explored theoretically and computationally over the past two decades \cite{balandin2008superior, cai2010thermal, lindsay2010flexural, wang2012tunable, wang2014phonon, lin2014thermal, wang2014two, nika2017phonons, chakraborty2018carbon,dai2019metal}.

Recently, there has been significant interest in novel applications of graphene in energy and environmental technologies. A critical aspect of these applications is the heat transfer between graphene and water, which often dictates overall performance. For instance, solar-thermal vapor generation systems utilizing graphene architectures are highly efficient in absorbing sunlight and converting it into heat \cite{wu2019multifunctional, wu2019scalable}. Efficient heat transfer between graphene and water is essential in these systems to enable rapid water heating and vaporization. Additionally, the laser or microwave manufacturing and processing of graphene structures, which involves creating carbon architectures with nanoscale features such as holes or pillars \cite{wang2020scalable, bi2022scalable, zhang2023review, wang2024manufacturing}, often employ solution-based techniques where graphene flakes are dispersed in water. In these processes, the heat transfer dynamics between the carbon structures and water are crucial in determining the thermal transients during manufacturing, which significantly influence the microstructure and performance of the final product. Thus, it is imperative to develop a comprehensive understanding of the thermal transport mechanisms between graphene structures and water and to formulate strategies to effectively control these thermal transport properties.

In solar-thermal vapor generation applications or laser manufacturing scenarios, graphene is rarely in its pristine form. More commonly, graphene contains defects and is at least partially oxidized. The primary oxygen-containing functional groups on graphene oxide (GO) nanosheets include hydroxyls (–OH), epoxides (C–O–C), carboxylic acids (–COOH), and other carbonyl groups (C=O). While pristine graphene is hydrophobic, these functional groups render GO hydrophilic by enhancing interactions with water molecules. It is widely accepted that hydroxyl and epoxy groups are primarily located on the basal plane, whereas carboxylic groups are predominantly found at the edges of GO nanosheets.

Recent experimental and computational studies have explored the effect of these oxygen-containing functional groups on thermal transport between graphene (and related structures) and water. A few molecular dynamics studies have specifically investigated the impact of functional groups on thermal transport between multilayer graphene (MLG) and water. Key findings include a significant increase in interfacial thermal conductance $G$ between MLG and water due to substantial hydrogen bonding between functional groups and water molecules \cite{li2022role}. Additionally, $G$ is highly dependent on the number of MLG layers, which modifies the phonon spectra of MLG \cite{alexeev2015kapitza, cao2018enhanced}. Despite these studies, there remains a limited understanding, particularly a quantitative analysis, of the effect of surface functional groups on thermal transport between single-layer graphene and water—an important consideration in the laser or microwave processing of graphene flakes in solutions.

In this work, we develop a deep neural network (DNN) interatomic potential for a system of graphene, GO, and water from ab initio molecular dynamics (AIMD) data. Leveraging this DNN potential, we perform molecular dynamics simulations to investigate the heat transfer between graphene at different oxidation levels with water during steady-state and transient heating conditions. 

\section{Methodology}
This study involves the development of a DNN interatomic potential derived from AIMD data. Subsequently, this potential is applied in transient molecular dynamics (MD) and nonequilibrium molecular dynamics (NEMD) simulations to assess the interfacial thermal conductance at the GO-water interface. The detailed methodology for each approach is outlined below.

\subsection{Development of Deep Neural Network Interatomic Potential}

Initially, AIMD simulations are performed utilizing the Vienna ab initio simulation package (VASP) \cite{kresse1993ab,kresse1996efficiency} for various GOs, carbon nanotube oxide (CNTO), GO-water, and CNTO-water configurations, as summarized in Table~\ref{tab:table1}. These simulations include GOs and CNTOs with diverse O/C ratios, ranging from 0 (pristine graphene or CNT) to 70\%, in both dry and aqueous environments. Temperatures ranging from 300 K to 1500 K and pressures from 0 GPa to 10 GPa are explored to capture atomic behavior across a broad range of thermodynamic conditions. Additionally, simulations involving GOs and CNTOs with free edges and holes are conducted to understand their interactions with water better.
\begin{figure}
\centering 
\includegraphics[width=0.35\textwidth]{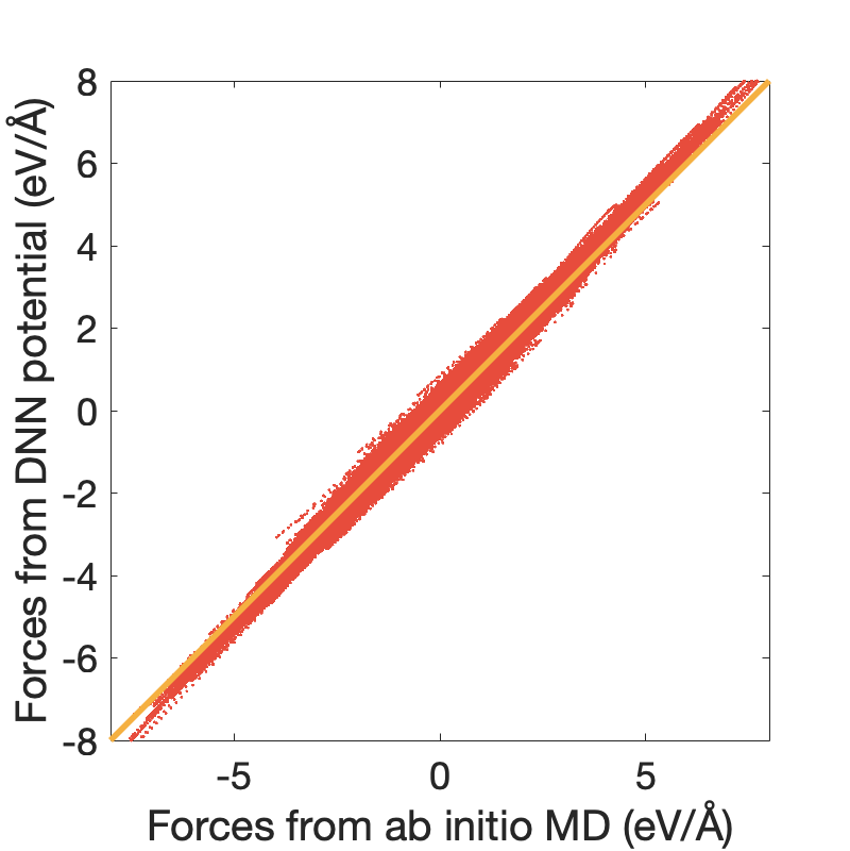}
\caption{Comparison of atomic forces predicted by AIMD and DNN potential.}
\label{fig:deepmd}
\end{figure}

\begin{table}
    \centering
    \begin{tabular}{lllll}
    Structure&Ensemble&Temperature (K)&Pressure (GPa)&Number of Data\\
    \hline
    Graphene & NVT & \begin{tabular}{l}200-400\\800\\300-1,000\end{tabular}& N/A & \begin{tabular}{l}5,931\\10,000\\10,000\end{tabular} \\
     \hline
    Graphene-water & NPT & \begin{tabular}{l}380-480\\300-1,000\end{tabular} & \begin{tabular}{l}0\\0.1\end{tabular} & \begin{tabular}{l}5,000\\5,000\end{tabular} \\
     \hline
    \begin{tabular}{l}GO(25\%O/C)\\of 8 configurations\end{tabular} & NVT & \begin{tabular}{l}300-400\\300-1,000\\1,000\\300-500\\1,500\end{tabular} & N/A & \begin{tabular}{l}38,081\\16,661\\7,060\\5,000\\8,747\end{tabular} \\
     \hline
    \begin{tabular}{l}GO(25\%O/C)-water\\of 8 configurations\end{tabular} & \begin{tabular}{l}NVT\\NPT\\NVT\\NVT\\NVT\\NPT\\NPT\\NPT\end{tabular} & \begin{tabular}{l}300-400\\300-400\\300-600\\500\\800\\380-480\\1,000\\800-300\end{tabular} & \begin{tabular}{l}N/A\\0.1\\N/A\\N/A\\N/A\\0\\0.1\\1.0\end{tabular} & \begin{tabular}{l}9,452\\2,046\\8,990\\7,042\\2,538\\1,752\\695\\3,103\end{tabular}\\
     \hline
    \begin{tabular}{l}5\%O/C CNTO-water\\D=$0.55$nm\end{tabular} & NPT & 900 & 0 & 5,000\\
 \hline
    \end{tabular}
    \caption{List of the structures and thermodynamic conditions for AIMD data used to train the DNN potential for this work.}
    \label{tab:table1}
\end{table}

The AIMD simulations adopt a $\Gamma$-centered scheme, utilizing a plane-wave cutoff energy of 600 eV, a Gaussian smearing width of 0.05 eV, and a convergence threshold of 10\textsuperscript{-8} eV for self-consistent electronic iteration to ensure accurate force calculations. Van der Waals interactions are accounted for using the Becke-Johnson damping function approach implemented in VASP. The resulting data, comprising instantaneous lattice vectors, atomic trajectories, atomic forces, and system energies, are leveraged to train a DNN interatomic potential via the DeePMD-kit package \cite{wang2018deepmd,zhang2020dp}.

The training process incorporates full relative coordinates to construct the descriptor, employing a cutoff radius of 6.0 \AA~ along with a smoothing cutoff radius of 0.5 \AA. The embedding network comprises three hidden layers with 25, 50, and 100 neurons, respectively, while the fitting network consists of three hidden layers, each containing 240 neurons. The initial learning rate, decay steps, and decay rate are set at 0.001, 16 million, and 3.51$\times 10^{-8}$, respectively. Figure \ref{fig:deepmd} compares the atomic forces predicted by these two approaches, again demonstrating strong agreement and providing additional confirmation of the accuracy of our DNN potential in approximating AIMD results.

\subsection{Transient molecular dynamics and lumped capacitance model}

The MD simulations in this study utilize the LAMMPS package \cite{thompson2022lammps}, augmented with the DeePMD plugin to harness the developed DNN potential. These simulations are executed on NVIDIA Tesla P100 GPUs. The periodic boundary condition is imposed along all three directions in the MD simulations, with a time step size of 0.1 fs. 

\begin{figure*}
\centering
\includegraphics[width=0.95\textwidth]{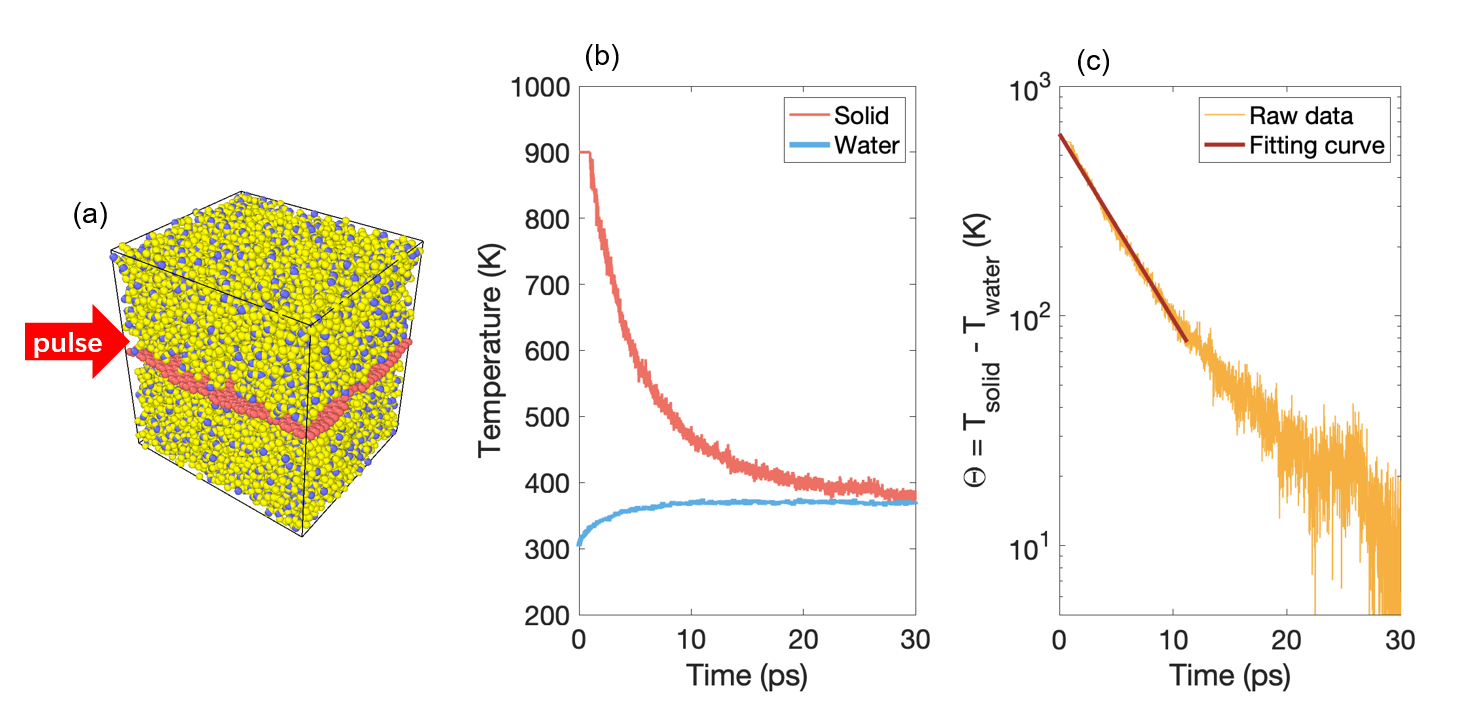}
\caption{Illustration of transient MD simulation. A short-period thermal pulse is added to (a) GO-water model. The time-dependent temperature decay curves of solid material and water are shown in panel (b) and their temperature difference is depicted in panel (c).}
\label{fig:transient}
\end{figure*}

Figure~\ref{fig:transient}a  displays the representative GO-water systems studied via transient MD simulations in this work. Initially, atoms are assigned random initial velocities corresponding to an average temperature of 5 K. To ensure full relaxation of the structures and the presence of water molecules in a liquid state, we employ four NPT stages by utilizing the Nos\'{e}-Hoover thermostat and barostat (isothermal-isobaric relaxation) \cite{nose1984unified,hoover1985canonical}. In the initial two NPT stages, the system is heated gradually from 5 K to 400 K over 50 ps and then maintained at 400 K for an additional 50 ps. Two subsequent NPT stages gradually cool down the system from 400 K to the target temperature within 50 ps. To mimic the transient heating of GO within water, immediately following the isothermal-isobaric relaxation of the system, we instantly elevate the temperature of the graphene layer to 900 K, using a temperature-rescaling thermostat implemented in LAMMPS, while allowing water molecules to freely evolve. This temperature is sustained for 1 ps, after which the thermostat is removed, enabling the entire system, including both GO and water, to evolve under the NVE ensemble. In Figure~\ref{fig:transient}b, we present representative temperature histories of GO and water from our transient MD simulations. These curves demonstrate the rapid decay of GO temperature and the corresponding rise in water temperature, reflecting the dissipation of heat from the hot solid to the cooler water medium.

The interfacial thermal conductance $G$ between GO and water can be determined by fitting an analytical solution to the temperature history obtained from transient MD simulations. This analytical solution is derived from the lumped capacitance assumption, which considers the energy balance in both the solid material (pristine graphene or GO in this work) and water. This approach yields two coupled equations:
\begin{subequations}
\begin{align}
-A_{sur}G(T_{s}-T_{w}) &= C_{s}V_{s}\frac{\partial T_{s}}{\partial t}  
\label{eqn:esolid}, \\
A_{sur}G(T_{s}-T_{w}) &= C_{w}V_{w}\frac{\partial T_{w}}{\partial t},  
\label{eqn:ewater} 
\end{align}
\end{subequations}
where subscripts $s$ and $w$ denote solid (graphene or GO) and water, respectively, and $A_{sur}$, $T$, $C$, $V$, and $t$ are surface area (including both sides), temperature, heat capacity, volume, and time, respectively. 

Subtracting Eq.~\ref{eqn:esolid} from Eq.~\ref{eqn:ewater} yields the differential equation:
\begin{equation}
\frac{d\Theta}{dt} = -\frac{\Theta}{\tau_{t}},
\label{eqn:ODE}
\end{equation}
where $\Theta$ is defined as the temperature difference between the solid material and water:
\begin{equation}
\Theta = T_{s}-T_{w},
\label{eqn:Theta}
\end{equation}
and $\tau_{t}$ is given by:
\begin{equation}
\tau_{t} = {\left(\frac{1}{C_{s}V_{s}}+\frac{1}{C_{w}V_{w}}\right)}^{-1}\frac{1}{A_{sur}G}.
\label{eqn:tau}
\end{equation}
The solution to Eq.~\ref{eqn:ODE} is straightforward:
\begin{equation}
\Theta(t) = \Theta_{i}\exp\left(-\frac{t}{\tau_{t}}\right),
\label{eqn:lump}
\end{equation}
where $\Theta_{i}=T_{s}(t=0)-T_{w}(t=0)$.

Clearly, Eq.~\ref{eqn:lump} demonstrates that the temperature difference between GO and water, denoted as $\Theta(t)$, exponentially decays over time following transient heating of the solid material.

In Figure~\ref{fig:transient}c, a representative series of $\Theta(t)$ data from our transient MD simulations is presented. Notably, the initial decay of the raw data exhibits exponential behavior (manifesting as a linear trend in the logarithmic scale), consistent with the prediction of Eq.~\ref{eqn:lump}. By fitting the data to Eq.~\ref{eqn:lump}, we can extract the value of $G$, representing the thermal conductance between GO and water.

\subsection{Nonequilibrium molecular dynamics}

\begin{figure*}
\centering
\includegraphics[width=0.65\textwidth]{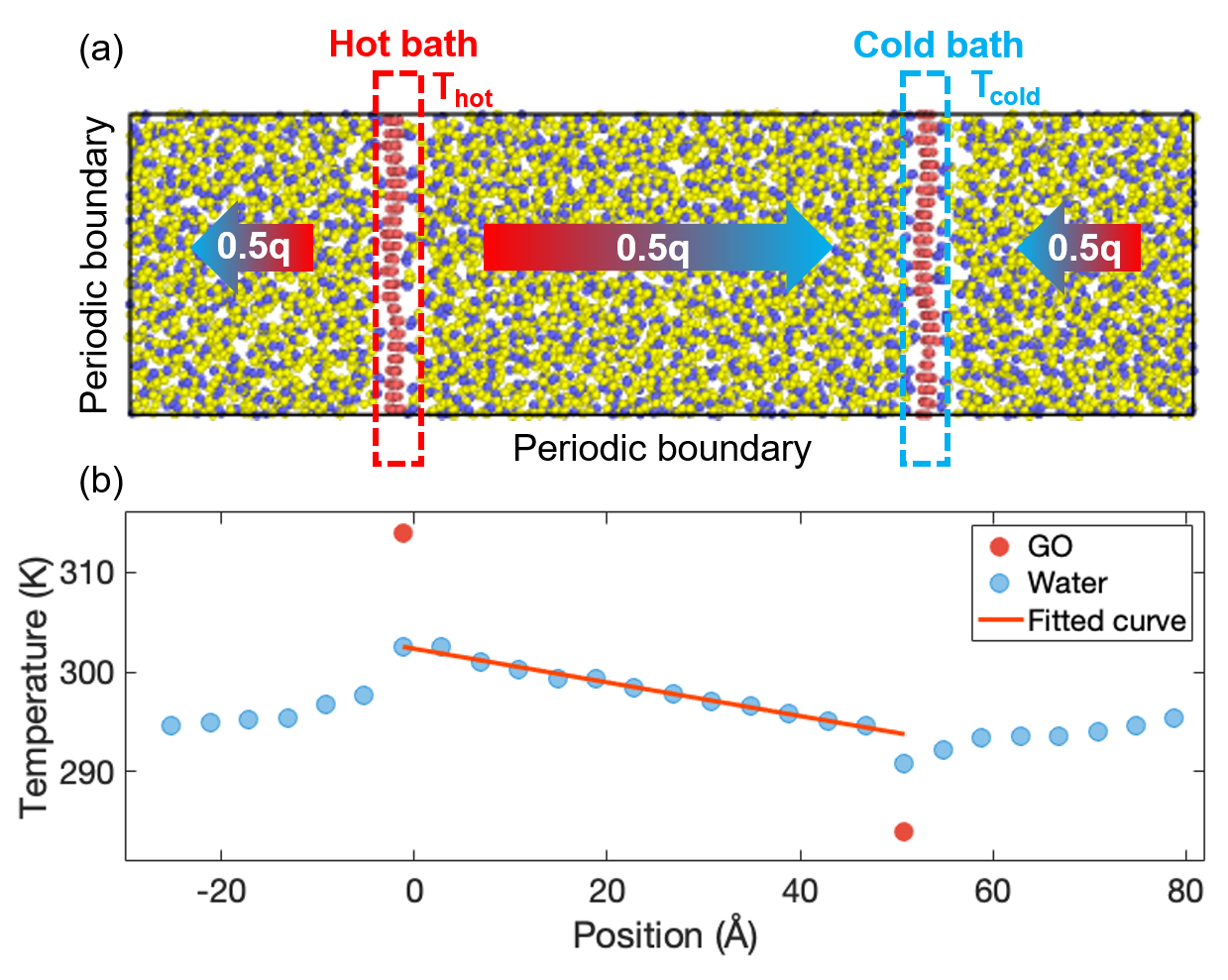}
\caption{Illustration of (a) NEMD simulation setup for GO-water model and (b) temperature profile obtained from steady-state. }
\label{fig:NEMDsetup}
\end{figure*}

As depicted in Fig.~\ref{fig:NEMDsetup}a, the NEMD simulations feature two GO layers serving as the hot and cold baths, respectively, inserted within the water at $0.25\times$ and $0.75\times$ the system length. In this standard NEMD setup, heat from the hot bath can transfer in either the positive direction or the negative direction (crossing the periodic boundary) to the cold bath. Since both heat pathways have the same length, i.e., $0.5\times$ the system length, the total heat transfer rate $q$ from the hot bath evenly divides into two $0.5q$ heat currents.

The cross-sectional area of the modeled system in NEMD simulations is $A_{c} =$30 \AA $\times$ 30 \AA. To ensure full relaxation of the structures and the presence of water molecules in a liquid state, we also employ four NPT stages as mentioned in the transient MD simulations with a time step size of 0.1 fs. Following four NPT stages, the simulation transitions to plain time integration in LAMMPS. Nos\'{e}-Hoover thermostats maintain the temperatures of the hot and cold baths at $1.05\times$ and $0.95\times$ the target temperature, respectively. Meanwhile, the NVE ensemble is applied to all other atoms in the system. This process extends for 1,000 ps to establish a steady-state heat flow from the hot bath to the cold bath. A representative steady-state temperature profile along the heat flow direction, obtained from our NEMD simulations, is shown in Fig.~\ref{fig:NEMDsetup}b.

Finally, the interfacial thermal conductance between GO and water can be estimated as:
\begin{equation}
G = \frac{q}{A_{c}\Delta T_{\text{interface}}},
\label{eqn:G_NEMD}
\end{equation}
where $\Delta T_{\text{interface}}$ represents the temperature difference between the heat bath and the temperature of water extrapolated to the position of the heat bath, as schematically illustrated in Fig.~\ref{fig:NEMDsetup}b. The value of $\Delta T_{\text{interface}}$ may slightly vary ($<$10\%) between the hot bath and the cold bath, with this difference utilized to estimate the error bar of the extracted $G$ values in this study.

\section{Results and Discussions}

\subsection{Interface structure of GO-water}
\begin{figure*}
\centering 
\includegraphics[width=0.7\textwidth]{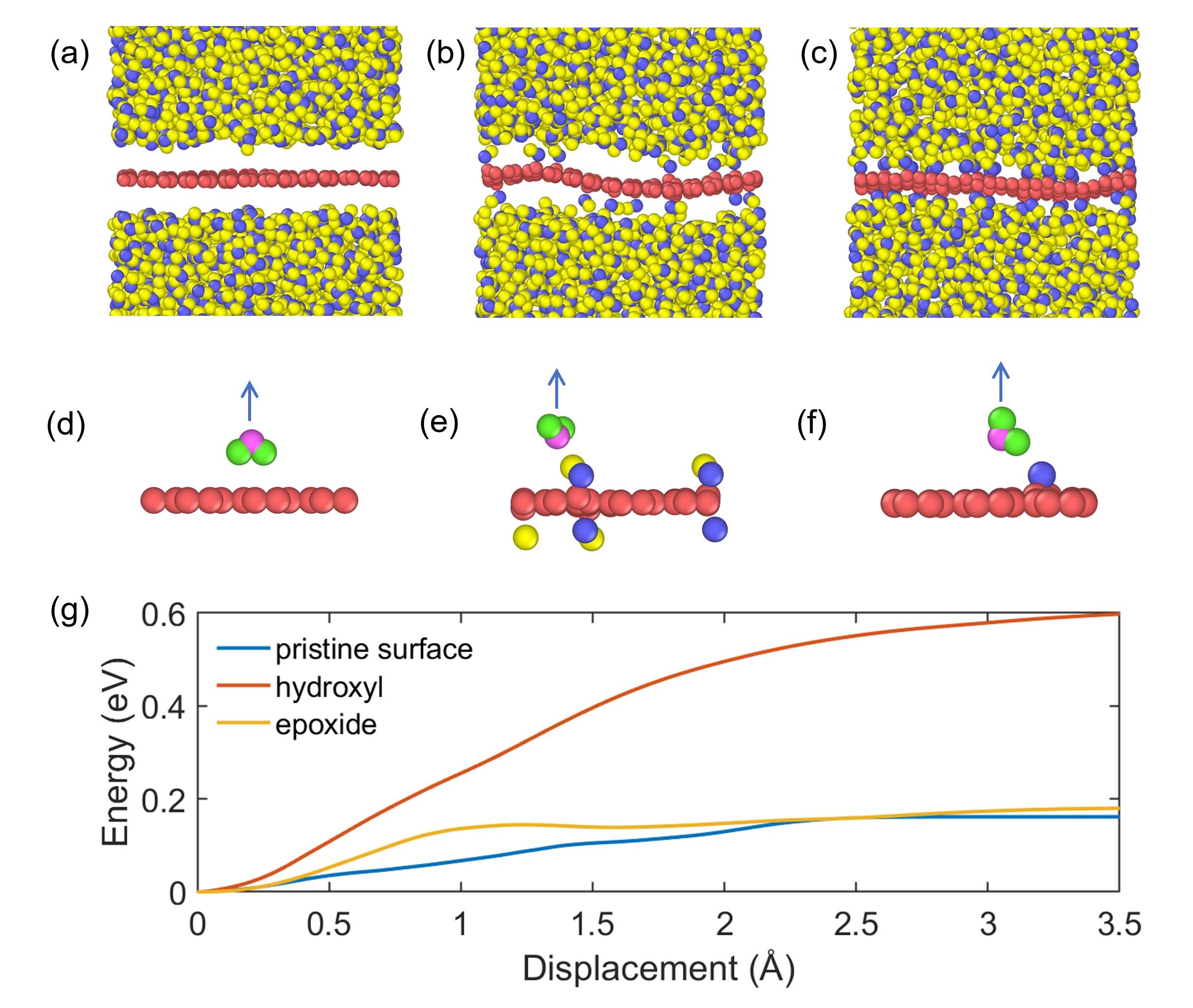}
\caption{Atomic structures of interfaces between graphene or GO and water for (a) pristine graphene-water model, (b) GO(0.05)-water model, and (c) GO(0.25)-water model. Illustrations of displacing a water molecule from its equilibrium position near (d) a pristine graphene surface, (e) a hydroxyl group, and (f) an epoxide group, where the blue arrow indicates the direction that the water molecule moves. (g) The relationship between displacement of the water molecule and its energy. }
\label{fig:structure}
\end{figure*}

Figure \ref{fig:structure} displays the snapshot of the relaxed interface between pristine graphene and water (panel a) and between GO and water (panels b and c for 5\% O/C and 25\% O/C, respectively) obtained from our MD simulations. As shown in Fig. \ref{fig:structure}a, there is a clear gap between pristine graphene surface and bulk water, which has a separation distance of 3.4 \AA. In contrast, water molecules can attach closely to the oxygen functional groups in the GO cases in Figs. \ref{fig:structure}b and c. 

To investigate the difference between the case of pristine graphene-water interaction and GO-water interaction, we performed additional simulations by moving a water molecule away from a pristine graphene surface, a hydroxyl group, or an epoxide group. As depicted in Figs.  \ref{fig:structure}d, e, and f, the simulations involve a small pristine graphene or a GO containing a few hydroxyl or epoxide groups and a single water molecule. The system is first minimized to achieve its lowest possible energy, allowing the water molecule to reach an equilibrium distance and relative position with the bare graphene surface or a functional group. We then gradually move the water molecule away from the graphene surface while monitoring the system's energy. As shown in Fig.  \ref{fig:structure}g, the energy of the hydroxyl group case increases significantly faster than that of the epoxide group case and pristine graphene surface, indicating a stronger interaction with water for hydroxyl groups than epoxide groups, while pristine graphene surface has the weakest interaction with water. As we will show later, this directly affects heat transfer from graphene or GO to water.  

\subsection{Temperature transients after pulsed laser heating}

\begin{figure*}
\centering 
\includegraphics[width=\textwidth]{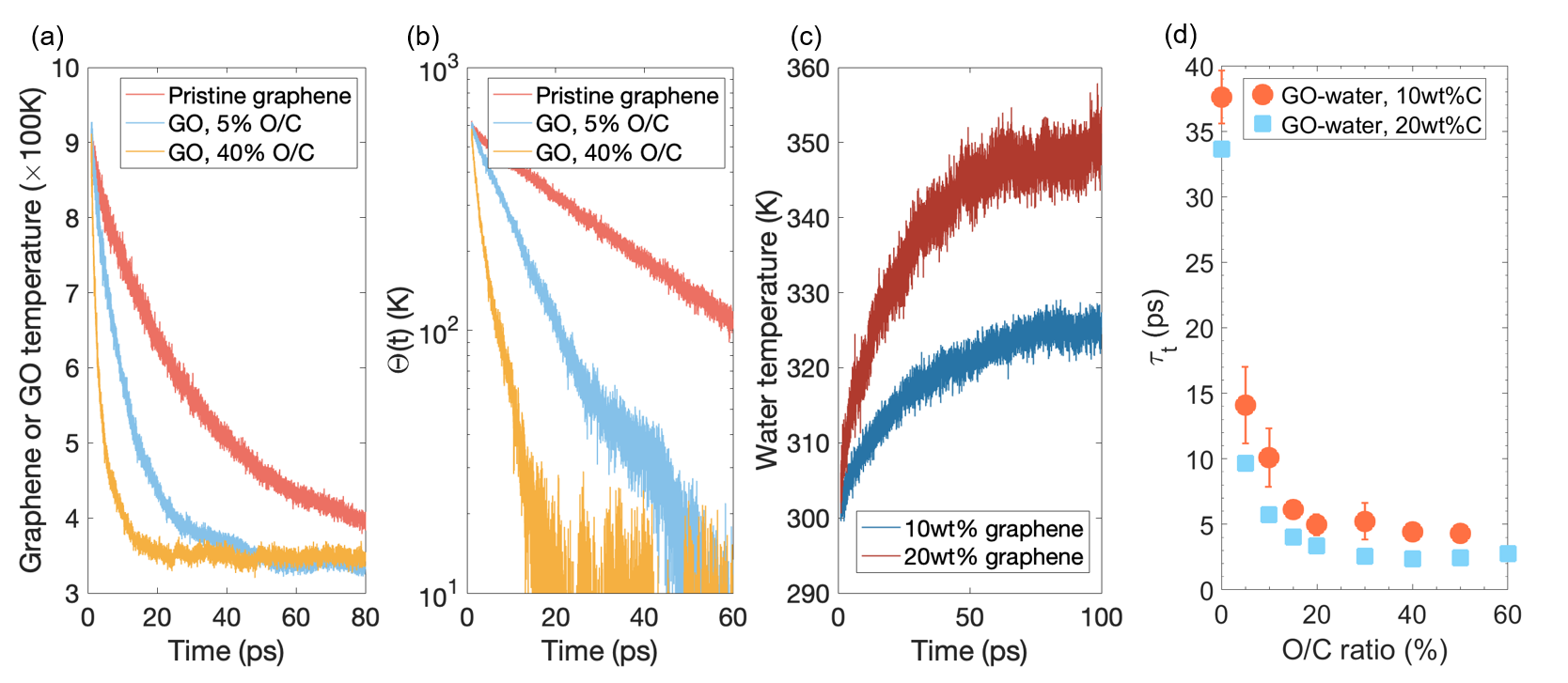}
\caption{Time-dependent (a) temperature of graphene or GO, (b) temperature difference between graphene or GO and water, and (c) water temperature of GO-water systems with different carbon weight percentage. (d) The O/C-ratio dependent thermal relaxation time ($\tau_{t}$) of GO-water with 10\% and 20\% carbon weight percentage.}
\label{fig:transient_temp}
\end{figure*}

Figure~\ref{fig:transient_temp} illustrates our transient MD simulation data for GO-water systems with 10 weight percentages of carbon (wt\%C) and 20 wt\%C. In Fig.~\ref{fig:transient_temp}a, immersed in water, pristine graphene exhibits a notably slower temperature decay compared to its oxidized counterparts. Specifically, it takes more than 80 ps for pristine graphene to cool down, whereas the temperatures of 5\%-O/C GO and 40\%-O/C GO have decayed to nearly room temperature within 40 ps and 20 ps, respectively. According to our lumped capacitance model's analytical solution in Eq.~\ref{eqn:Theta}, the difference between the GO temperature and the water temperature, denoted as $\Theta$ (Eq.~\ref{eqn:Theta}), should decay exponentially with a time scale of $\tau_{t}$ (Eq.~\ref{eqn:tau}). As demonstrated in Fig.~\ref{fig:transient_temp}b, the $\Theta$ of pristine graphene, 5\%-O/C GO, and 40\%-O/C GO all exhibit well-defined exponential decay behavior, enabling us to extract the thermal relaxation time $\tau_{t}$.

In Fig.~\ref{fig:transient_temp}d, the $\tau_{t}$ of GO-water systems of 10 wt\%C and 20 wt\%C for various O/C ratios of GO are depicted. A higher O/C ratio of GO substantially accelerates heat dissipation into water. For instance, the $\tau_{t}$ of pristine graphene in water (10 wt\%C) is 37.6 ps, whereas it is only 4.4 ps for the 40\%-O/C GO, indicating a one-order-of-magnitude acceleration.

The carbon weight percentage of GO in water does not significantly impact $\tau_{t}$, although we anticipate observing a more pronounced difference at higher (but less frequently used in practical applications) weight percentages. Nevertheless, it directly influences the final temperature of water; as anticipated, the water temperature is higher when the system has a higher weight percentage of laser-absorbing solids, as shown in Fig.~\ref{fig:transient_temp}c. 

\subsection{Interfacial thermal transport}

\begin{figure*}
\centering 
\includegraphics[width=0.8\textwidth]{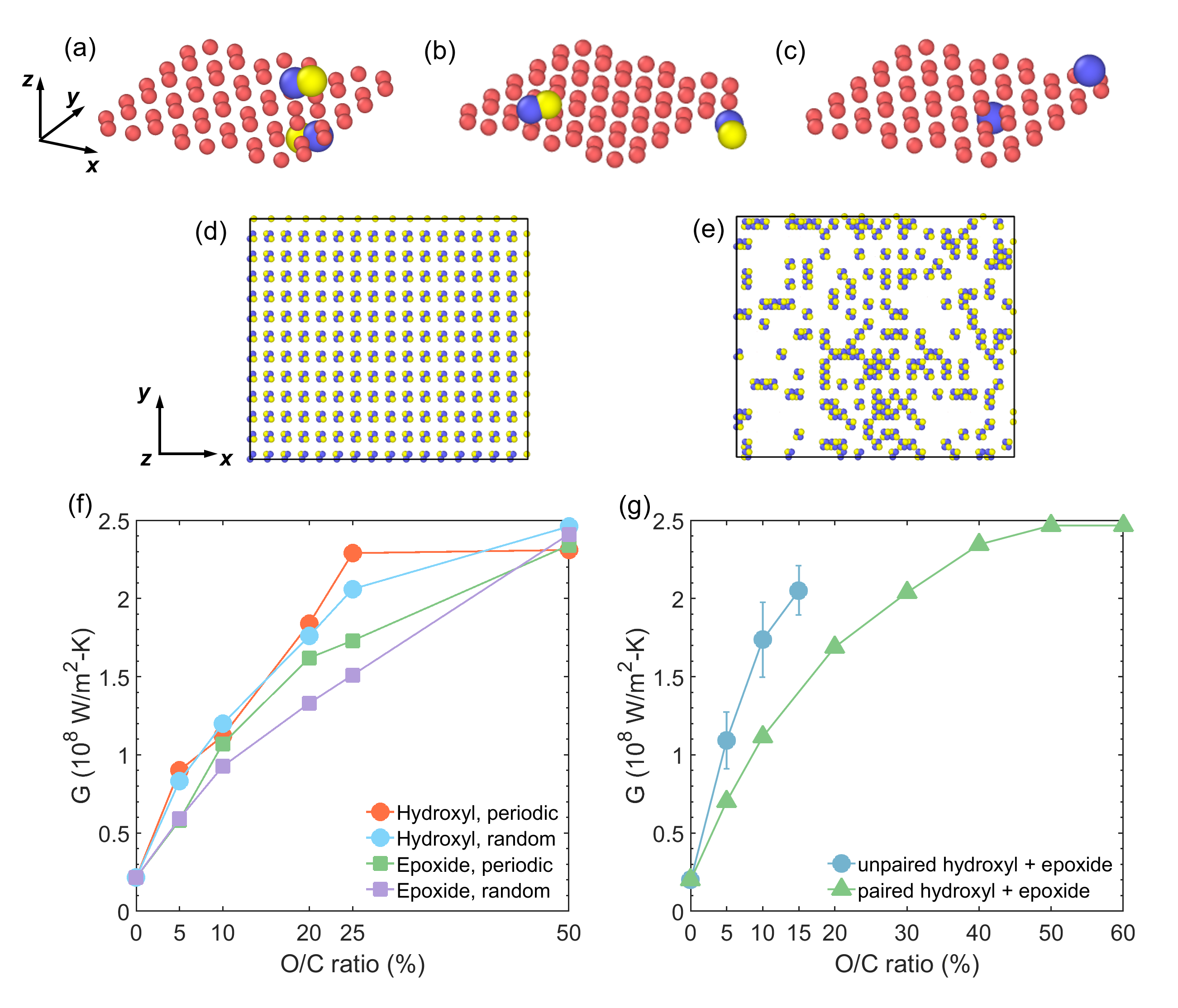}
\caption{Atomic structures of GOs with (a) paired hydroxyl groups, (b) unpaired hydroxyl groups, and (c) epoxide groups. Top view of (d) periodically distributed hydroxyl groups and (e) randomly distributed hydroxyl groups, where the carbon atoms are hidden. O/C-ratio-dependent $G$ of GO obtained from (f) transient simulations and (g) NEMD simulations.}
\label{fig:G}
\end{figure*}

Several representative GO structures are investigated in this work. As illustrated in Fig. \ref{fig:G}, we investigate GO with hydroxyl groups and with epoxide groups, which are the two most common oxygen-containing functional groups on the surface of graphene. Notably, hydroxyl groups are most stable when they form in pairs (as shown in Fig. \ref{fig:G}a), i.e., the two hydroxyl groups are bonded to two neighboring carbon atoms but located on opposite sides of the graphene layer. Unless otherwise noted, this paired configuration is used for hydroxyl functionalization in our simulations. To also consider the less stable configuration of unpaired hydroxyl groups, we simulate GO with unpaired hydroxyl groups (as shown in Fig. \ref{fig:G}b). The epoxide groups are simply located on top of the center of two neighboring carbon atoms, as shown in Fig. \ref{fig:G}c. Other functional groups, such as carboxyl and carbonyl, are not considered as they are more often found at graphene edges. In addition to different types of functional groups, we also investigate periodic versus random (closer to the real-world scenario) arrangements of functional groups on the graphene surface, as depicted in Figs. \ref{fig:G}d and e, respectively.

Figures \ref{fig:G}f and g present the interfacial thermal conductance between GO and water as a function of the O/C ratio of GO, as predicted by transient MD simulations (panel f) and NEMD simulations (panel g), which show an overall agreement between the two approaches. There are multiple remarkable observations, as discussed as follows. 

Firstly, the thermal conductance $G$ of the GO-water interfaces exhibits nearly linear growth with the O/C ratio in low to moderately oxidized GO (i.e., O/C ratio $<40\%$), plateauing beyond 40\%. Specifically, $G$ is $0.21\times10^8$ W/m$^{2}$-K for pristine graphene (i.e., 0\% O/C ratio), while it reaches $2.35\times10^{8}$ W/m$^{2}$-K for 40\%-O/C-ratio GO, representing a remarkable one-order-of-magnitude increase in interfacial thermal transport capability. Notably, oxidizing pristine graphene to a 5\% O/C ratio can increase its $G$ with water by 248\%, indicating a significant impact of oxygen-containing functional groups on the basal plane of graphene on thermal transport between graphene and water. The significant increase in $G$ by functionalization is not surprising, because, as shown in Fig. \ref{fig:structure}g, hydroxyl groups and epoxide groups have stronger interaction with water than a pristine graphene surface.  

\begin{figure*}
\centering 
\includegraphics[width=0.5\textwidth]{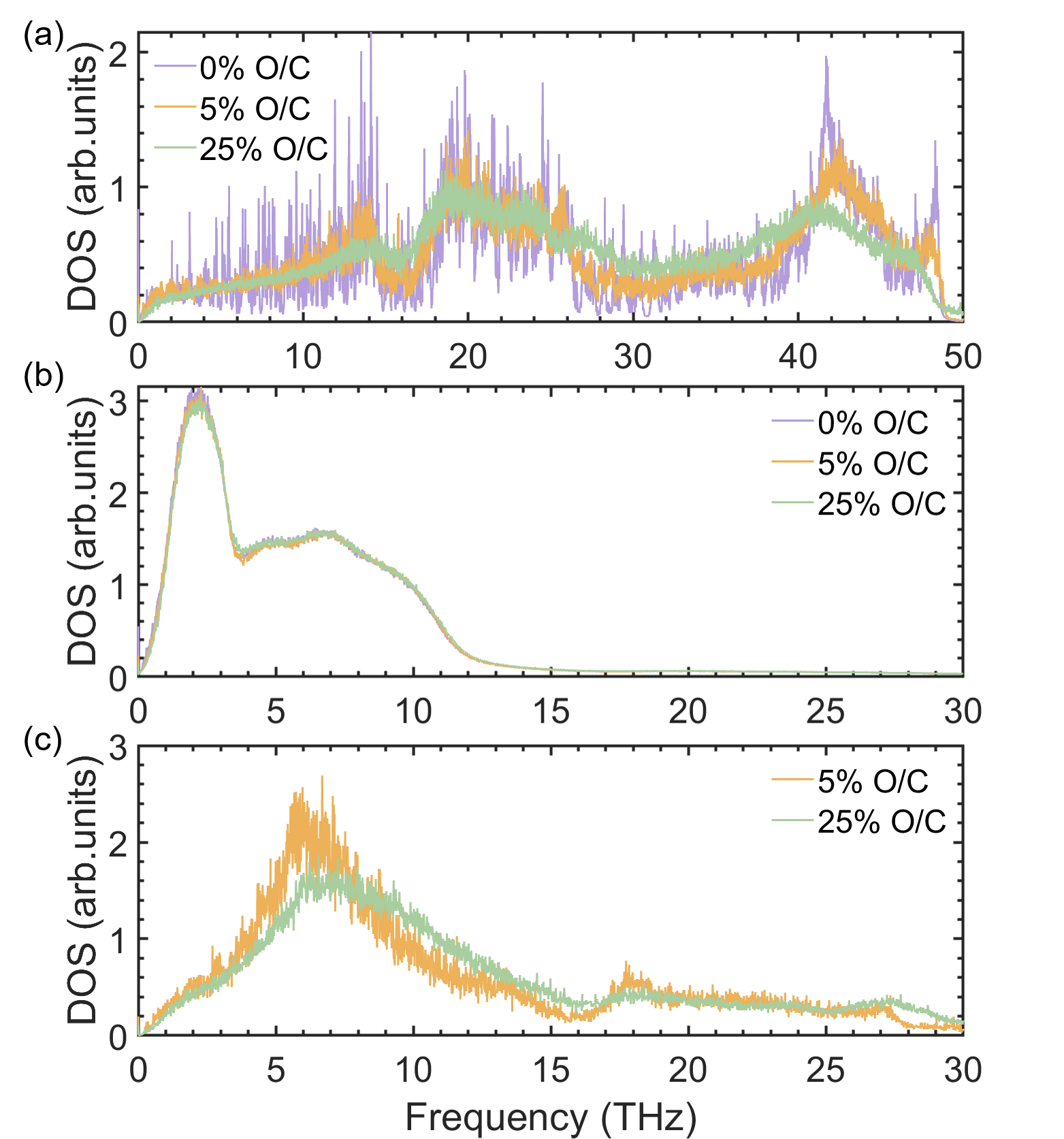}
\caption{Phonon vibrational density of states of (a) carbon atoms, (b) oxygen atoms belonging to water molecules, and (c) oxygen atoms of functional groups obtained from GO-water models with 0\%, 5\%, and 25\% O/C ratios.}
\label{fig:DOS}
\end{figure*}

In addition to the stronger interactions between functional groups and water molecules, we note that the functional groups have a vibrational spectrum that largely overlaps with that of water molecules. As shown in Fig. \ref{fig:DOS}, the vibrational density of states (vDOS) of carbon atoms (panel a) in the graphene layer spans a wide frequency range up to 50 THz, while the vDOS of oxygen atoms in water molecules (panel b) and those in functional groups (panel c) are much narrower. Notably, the vDOS of the latter two exhibit significant overlap, which, similar to other materials, often leads to efficient heat transfer between them.

\begin{figure*}
\centering 
\includegraphics[width=1.0\textwidth]{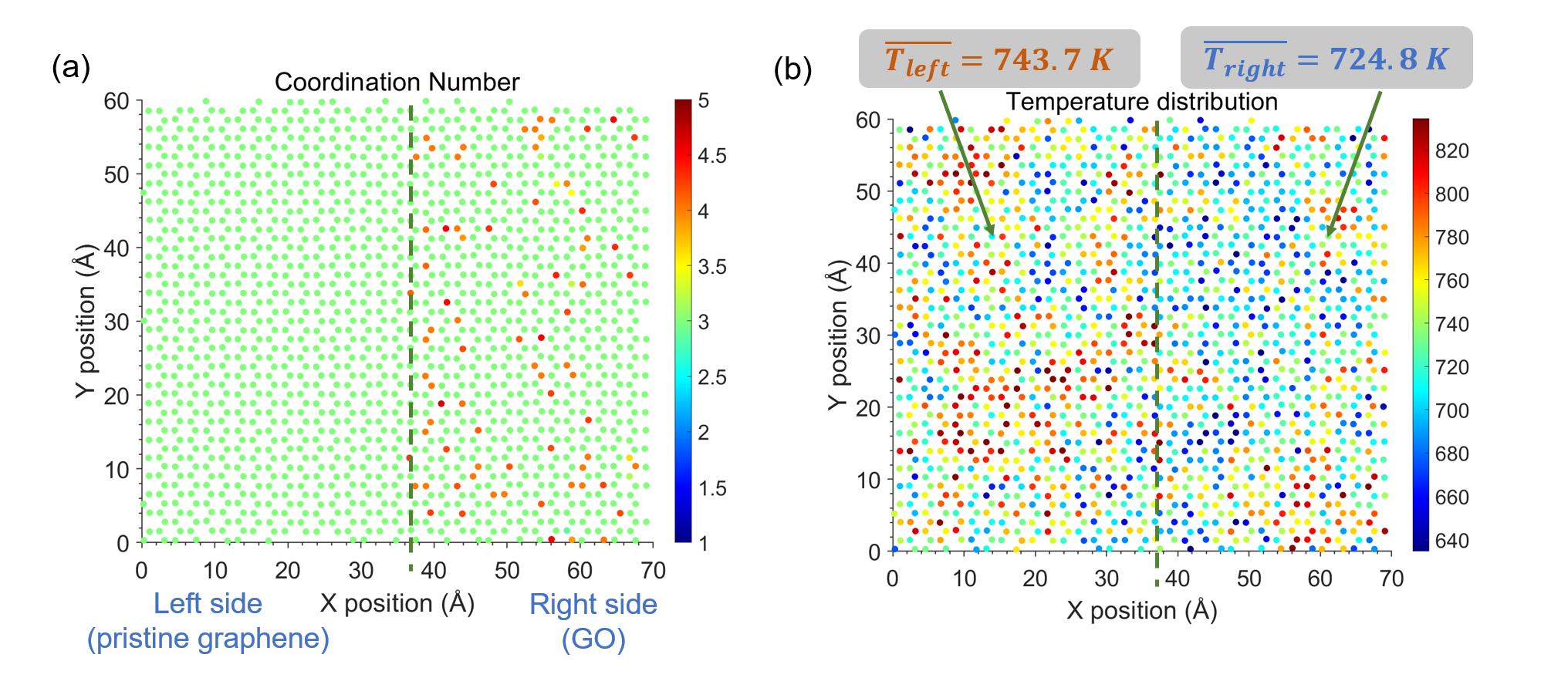}
\caption{Illustration of coordination number and temperature contour of GO layer obtained in the transient simulation of GO-water. (a) Coordination number of GO layer. (b) Temperature contour of GO layer.}
\label{fig:Tcontour}
\end{figure*}

To further confirm the effect of functional groups on heat dissipation from graphene to water, we apply transient MD simulations to study the thermal transport process between water and a half-pristine/half-functionalized GO structure, as illustrated in Fig. \ref{fig:Tcontour}a. Specifically, the left half of the GO is clean, free of functional groups or other defects, while the right half is functionalized, corresponding to an overall O/C ratio of 5\% for the entire GO structure. Initially, the graphene layer, submerged in water at 300 K, is heated to 900 K. After the heat source is removed, the temperature of the graphene continuously decreases until it reaches equilibrium with water, as already shown in Fig. \ref{fig:transient}b. During this process, we monitor the temperature of all the carbon atoms. As depicted in Fig. \ref{fig:Tcontour}b, the averaged temperature of the pristine half of the GO, which is recorded over a time period of 4 ps after removing the transient heat source, is notably higher than the functionalized half, indicating more efficient heat dissipation to water by the functionalized half. This further confirms that the functional groups enhance heat transfer to water compared to a pristine graphene surface.

Secondly, despite the inefficiency of interfacial thermal transport between pristine graphene and water than the case of GO, the value of $G=2.5\times10^{8}$ W/m$^{2}$-K for 50\%-O/C-ratio GO is comparable to MD predictions of interfacial thermal conductance $G$ of various nonmetallic solid-solid interfaces. Examples include CNT-copper \cite{wang2012two}, silicon-copper \cite{wang2012two}, silicon-germanium \cite{landry2009thermal}, and conceptual Lennard-Jones solid-argon systems \cite{wang2014decomposition}, which are in the range of $2-6\times10^{8}$ W/m$^{2}$-K. The experimentally measured values of $G$ of solid-solid interfaces \cite{hopkins2013thermal} are usually much lower than this range, because of the incomplete contact between solids (i.e., the existence of interface voids). However, considering that liquid water can well conform to the GO surface, it is expected that the actual $G$ of GO-water can be significantly higher than that of solid-solid interfaces.

Third, the arrangement of functional groups on the graphene layer, whether random or periodic, does not significantly affect the thermal conductivity ($G$), particularly at low O/C ratios. However, the impact becomes more pronounced at higher O/C ratios of 20\% and 25\%. In random structures, functional groups are closer to each other, which reduces heat transfer efficiency from the graphene layer to water compared to when the functional groups are more uniformly distributed. This phenomenon is analogous to macroscopic heat transfer scenarios, where periodically arranged fins on a base are generally more efficient than randomly arranged fins. The latter often leads to aggregation (not chemical), causing competition among the fins in conducting heat away from the hot base. This effect is demonstrated through finite element analysis simulations provided in the Supplementary Materials (Fig. S1). Additionally, functional groups within a cutoff distance of approximately 2.5-3 \AA$\:$ from each other can directly interact, potentially hindering thermal transport in the graphene-functional group-water channels. The interactions between functional groups at different distances are provided in the Supplementary Materials (Fig. S2), where inter-functional-group forces are evaluated. This is similar to how chain-chain interactions impede thermal transport within individual polymer chains in bulk polymers \cite{Henry2008,Shen2010,huang_2017,Ma_PVDF}.

At a 50\% O/C ratio, the $G$ values for both periodic and random structures become similar. This is because, in GO with a 50\% O/C ratio, the periodic structure is also crowded with functional groups, diminishing any advantage over the random arrangement.

Fourth, as shown in Fig. \ref{fig:G}f, the interfacial thermal conductance $G$ for GO with hydroxyl groups (circles) is significantly higher than that for GO with epoxide groups (squares). As displayed in Fig. \ref{fig:structure}g, the interaction between hydroxyl groups and water is much stronger than that between epoxide groups and water. This is why GO with hydroxyl groups exhibits higher thermal conductance to water than GO with epoxide groups at the same O/C ratio.

Lastly, it is significant to note that while $G$ increases rapidly with the O/C ratio at low O/C ratios, it appears to level off around 50\%. This trend is more clearly illustrated in Fig. \ref{fig:G}g, which presents the $G$ values from NEMD simulations with smaller uncertainty than the transient MD data. As shown, $G$ increases nearly linearly with the O/C ratio between 0\% and 20\%, but the rate of increase significantly slows down thereafter. A maximum O/C ratio of 60\% is investigated in this work, as higher ratios tend to cause significant spontaneous dehydration, leading to the detachment of functional groups from the graphene surface and the formation of water molecules. The leveling off of the $G$-O/C ratio curve can again be understood through an analogy with macroscopic fins. At higher O/C ratios, the functional groups (like fins) are too close to each other, competing in conducting heat away from the base, which has a finite thermal conductivity.

\section{Conclusion}
In conclusion, we developed a deep neural network interatomic potential for GO-water interactions based on ab initio molecular dynamics simulation data. Utilizing this potential, we modeled the thermal transport between single-layer GO and water through transient MD simulations, which mimic the laser heating of graphene in water, as well as NEMD simulations. Our results reveal a nearly one-order-of-magnitude increase in the interfacial thermal conductance $G$ between graphene and water with the addition of functional groups (hydroxyl and epoxide) to the graphene surface. Among these, hydroxyl groups significantly enhance $G$ more than epoxide groups, attributed to the stronger interaction between water and hydroxyl groups compared to epoxide groups. Furthermore, we observed that increasing the O/C ratio to 50\% and above does not lead to further improvement in $G$. This saturation effect occurs because, at high O/C ratios, the functional groups become too densely packed, leading to competition among them in conducting heat away from the hot graphene layer. The insights gained from this study are crucial for designing applications that rely on efficient heat transfer between graphene and water. Examples include laser manufacturing of graphene in solutions, water vaporization, and desalination applications where graphene is used to absorb sunlight and heat water.

\section*{Acknowledgements}
This study is partially supported by NSF Grants CMMI-1826392, CMMI-1825576, and CMMI-1726792. Cui, Rajan, and Wang also extend their thanks to the National Science Foundation EPSCoR Research Infrastructure Program (OIA-2033424). Additionally, the authors would like to acknowledge the support provided by the Research and Innovation team and the Cyberinfrastructure Team in the Office of Information Technology at the University of Nevada, Reno, for facilitating access to the Pronghorn High-Performance Computing Cluster. 

\appendix

\bibliographystyle{elsarticle-num} 
\bibliography{references}






\end{document}